# On the Convergence of Artificial Intelligence and Distributed Ledger Technology: A Scoping Review and Future Research Agenda
# (Working Paper)

Konstantin D. Pandl
Scott Thiebes
Manuel Schmidt-Kraepelin
Ali Sunyaev

Author email addresses:
{konstantin.pandl, scott.thiebes, manuel.schmidt-kraepelin, sunyaev}@kit.edu

Institute for Applied Informatics and Formal Description Methods
Karlsruhe Institute of Technology, 76131 Karlsruhe, Baden-Württemberg, Germany

**ABSTRACT** Developments in Artificial Intelligence (AI) and Distributed Ledger Technology (DLT) currently lead to lively debates in academia and practice. AI processes data to perform tasks that were previously thought possible only for humans. DLT has the potential to create consensus over data among a group of participants in uncertain environments. In recent research, both technologies are used in similar and even the same systems. Examples include the design of secure distributed ledgers or the creation of allied learning systems distributed across multiple nodes. This can lead to technological convergence, which in the past, has paved the way for major innovations in information technology. Previous work highlights several potential benefits of the convergence of AI and DLT but only provides a limited theoretical framework to describe upcoming real-world integration cases of both technologies. We aim to contribute by conducting a systematic literature review on previous work and providing rigorously derived future research opportunities. This work helps researchers active in AI or DLT to overcome current limitations in their field, and practitioners to develop systems along with the convergence of both technologies.

**KEYWORDS** Artificial intelligence, blockchain, distributed ledger technology, machine learning, technological convergence

I. INTRODUCTION

Artificial Intelligence (AI) and Distributed Ledger Technology (DLT) are among today's most actively debated developments in information technology with potential for tremendous impact on individuals, organizations, and societies over the next decades. A 2018 report from the McKinsey Global Institute estimates that the application of AI in various industries could deliver an additional global economic output of around USD 13 trillion by 2030 [1]. Similarly, a study by the World Economic Forum predicts that by 2025, up to 10 % of the world's GDP may be stored on a blockchain [2], which is the most commonly used concept of DLT today [3].

AI can perform complex tasks that were previously thought possible only for humans to perform. In some application domains, AI can nowadays already exceed human capabilities. Research in the health care domain has, for instance, shown that AI can analyze echocardiograms faster and more accurate than medical professionals [4]. Furthermore, advancements in AI are also expected to be key enablers of important upcoming innovations such as autonomous driving [5] or intelligent robots [6], to name but a few. DLT, on the other hand, can create consensus in a network of untrusting parties. It became known to the public due to the emergence of the cryptocurrency Bitcoin [7]. Following the success of Bitcoin, further DLT applications are emerging in application domains beyond finance that are often corresponding to those of AI. DLT may, for example, be used to manage access control for electronic health records [8], or to secure the IT systems of autonomous cars [9].

As one result of these developments, we now increasingly see the emergence of applications using both information technologies in close integration. Recent work, for example, uses deep reinforcement learning to explore attacks on blockchain incentive mechanisms [10]. In doing so, the integrated AI can detect new attack strategies and provide security insights, even for well-studied DLT protocols like Bitcoin. Another recent work uses a DLT-based platform to exchange data and computing resources in order to enable AI applications [11]. The platform gives data providers the opportunity to share their data while keeping it confidential and maintaining the right to manage data access. Data consumers can then train AI algorithms on the provided data and compensate data providers for use of their data.

The examples above demonstrate that the integration of AI and DLT yields great potential to advance the capabilities of both technologies, and, ultimately, to increase the positive impact of AI and DLT on individuals, organizations, and societies. Yet, in order to make meaningful contributions, researchers and practitioners alike will have to keep up with the latest developments in both fields as well as the most recent developments and innovations related to their integration. Owing to the fast pace and interdisciplinary nature of both research areas, assessing the current state of research on AI and DLT and especially their integration in its entirety is a difficult task. In attempt to provide guidance to researchers enticed by the integration of AI and DLT, previous research has either focused on partial aspects of the integration of AI and DLT such as the use of blockchain for AI (e.g., [12]) or on deriving conceptual ideas of how both technologies might be integrated with each other (e.g., [13]). Despite the invaluable contributions that these publications have made to the nascent stream of literature concerning the integration of AI and DLT, we currently lack in-depth knowledge about the current state of research on the integration of AI and DLT that does (a) not only focus on a specific DLT concept (i.e., blockchain), (b) considers the reciprocal integration of both technologies (as opposed to the one-way integration of, for example, DLT into AI), and (c) goes beyond a purely conceptual level. In particular, we still lack a comprehensive overview of the most pressing research challenges that must be overcome in order to unleash the full potential of integrating both technologies. With this research, we aim to address this apparent knowledge gap by asking and answering the following two research questions:

*RQ 1: What is the current state of research on the technological integration of AI and DLT?*
*RQ 2: What are open research challenges on the technological integration of AI and DLT?*

To address our research questions, we draw on the concept of convergence (see section II.C) and conduct a systematic literature review on the current state of research on the convergence (i.e., integration) of AI and DLT and develop a future research agenda. The contribution of our work is thereby four-fold. First, in contrast to extant research in this area, we also include non-blockchain distributed ledgers in our analysis of extant literature. Prior work has highlighted several current shortcomings of blockchain in the application for AI [12]. Other DLT concepts (e.g., BlockDAG or TDAG), may be more promising for solving some of these shortcomings [3] and thus potentially better suited for certain AI applications. Second, we consider the reciprocal nature of convergence and investigate both perspectives: the usage of AI for DLT, and the usage of DLT for AI. To the best of our knowledge, no holistic review on the convergence of AI and DLT exists today, which considers the large variety of interaction cases from both perspectives. Third, we aim to bridge the gap between theory and practice by drawing theoretical conclusions from practical research, as well as outlining future potential for practical and theoretical research from the theory in these fields. Fourth, we apply the concept of convergence [14, 15] as a theoretical lens for our article. In doing so, we contribute to the understanding of how convergence may drive product innovations and create economic value in the information technology (IT) industry. In addition, we demonstrate how convergence can be applied as a lens to tackle research questions in interdisciplinary, innovative, and emerging research fields.

The remainder of this article is organized as follows: In section 2, we discuss related work on AI and DLT, and introduce the concept of convergence. Afterward, we describe our methods in section 3. In section 4, we analyze the current literature on AI usage for DLT, and in chapter 5, provide our future research agenda on AI for DLT. In section 6, we analyze the current literature on DLT usage for AI, and a corresponding research agenda in section 7. In section 8, we discuss our results, before we conclude this article in section 9.

## II. RELATED WORK

### A. ARTIFICIAL INGELLIGENCE

AI enables computers to execute tasks that are easy for people to perform but difficult to describe formally. Such tasks typically occur in complex or uncertain environments [16]. Not only science fiction, but also ongoing debates in society discuss *Artificial General Intelligence*. The concept behind Artificial General Intelligence are computer programs that can control themselves and solve tasks in a variety of different domains [17]. In contrast, most AI-based deployed systems solve tasks in narrow application domains, and are referred to as *Narrow Artificial Intelligence*. Several approaches to design such narrow AI-based systems exist. For example, knowledge bases have seen a lot of attention by researchers in the past, but nowadays, Machine Learning (ML) seems to be the most well-spread approach toward building AI-based systems [16]. ML-based systems consist of a model that represents a function between input data and output data. In most cases, ML models have to be trained. In this training phase, an optimization algorithm tweaks the model parameters in order to minimize a loss or maximize a reward. Depending on the application, different types of training exist. In the case of supervised machine learning, the input data and the corresponding output data are known during the training phase. In the case of unsupervised machine learning, only the input data is known but no output data. In a reinforcement learning setting, a learning agent executes actions that result in an immediate reward, whereas the agent's goal is to maximize a future, cumulative reward. In general, the training phase can require large amounts of data and, thus, is often computationally intensive. This is especially the case for deep neural networks, which are complex ML models with many parameters that have paved the way for many of the recent advancements in ML [18]. In Figure 1, we present a high-level overview of different types of approaches toward designing AI-based systems.

The execution of a (trained) ML model is called inference. It is usually computationally less expensive than the initial training phase. Some models can only be described as a black box, meaning that its inner functionalities are difficult to explain. Among many research streams, some cutting-edge research in ML aims to better explain the inner functioning of ML models in order to guarantee their robustness [19]. Other streams aim to increase ML systems' capabilities [20, 21], or aim to ensure training data privacy when creating ML-based systems [16]. Besides these ML approaches introduced above, many variations exist. For example, for some ML algorithms, there is no explicit model building phase at all [22]. For a detailed overview of AI and ML, in particular, we refer to Russell and Norvig [23].

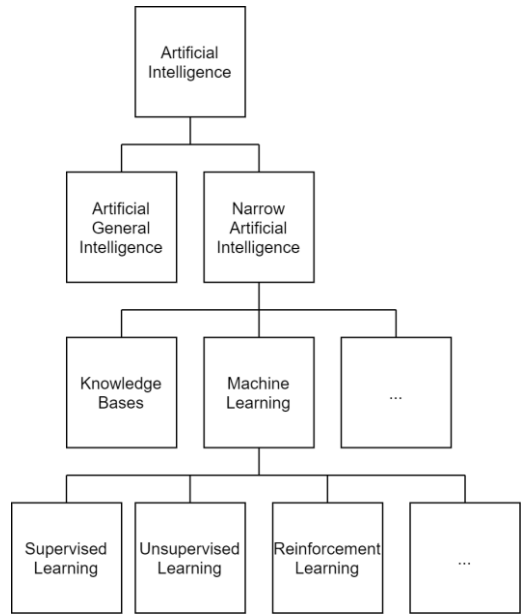

**FIGURE 1.** Hierarchical overview of Artificial Intelligence.

## B. DISTRIBUTED LEDGER TECHNOLOGY

DLT enables the operation of a highly available, append-only, peer-to-peer database (ie the distributed ledger) in untrustworthy environments characterized by Byzantine failures, where separated storage devices (ie the nodes) maintain a local replication of the data stored on the ledger. The most widespread concept of DLT, blockchain, became known to the public due to the emergence of the cryptocurrency Bitcoin [7]. Nowadays, applications beyond financial transactions are emerging, such as the management of medical data [24], the organization of robotic swarm systems [25], or decentralized games [26].

Today, several concepts for DLT exist with different characteristics, for example, regarding the transaction throughput or the fault tolerance. The most widespread concept of DLT is blockchain [3]. Other concepts, for example, rely on directed acyclic graphs [3, 27, 28]. We provide an overview of DLT concepts with different characteristics in Figure 2.

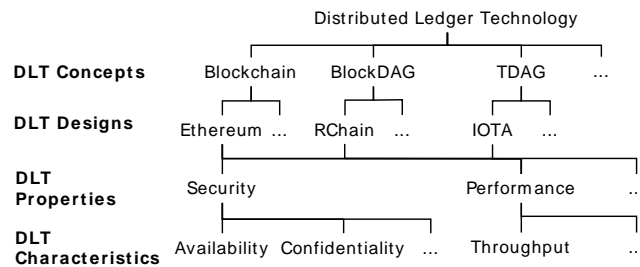

**FIGURE 2.** Hierarchical overview of Distributed Ledger Technology and its characteristics [3].

In a blockchain system such as Bitcoin [7], transactions of network participants are stored in blocks of a fixed size. Every block also includes a timestamp, and a hash of the previous block. Thus, the system can be regarded as a chain of blocks. Cryptographic techniques are used to ensure that only legitimate participants holding a cryptographic key can perform transactions, which are stored in the block. Bitcoin [7] was the first blockchain created and has mostly financial applications.

Bitcoin or Ethereum [29] specifically are unpermissioned blockchains, meaning that anyone can join the network with unrestricted access. To secure the network, only a selected node can propose a new block that includes the cryptographically signed transactions. This node has to find a block candidate with a hash below a certain, network-defined threshold. As this hash calculation is impossible to reverse, a node has to perform large computational efforts to find that new block, competing against other nodes. In return, the node that successfully found a block gets a reward in cryptocurrency payment. Since there are at any time many miners aiming to find the next block, the chances of finding the next block are relatively low for an individual miner. As a result, the variance of the mining payoff for an individual miner is relatively large. Therefore, most miners nowadays mine together in a mining pool. If the collective pool finds a block, the pool gets the block reward and distributes it among its miners according to their share of hash calculations. This reduces the payoff variance for individual miners participating in the pool. Besides Bitcoin, several implementations of blockchains exist with different characteristics [3]. Some implementations provide high transaction confidentiality guarantees (e.g., Zcash [30]), or enable a universally usable, decentralized Turing complete computing platform (e.g., Ethereum [29]). In the latter case, programs can be stored and executed upon the blockchain system. These programs are referred to as smart contracts.

## C. CONVERGENCE OF AI AND DLT

First described by Nathan Rosenberg in the 1960s, convergence describes a phenomenon in which two or more initially separate items move toward unity and become increasingly integrated with each other. Convergence typically occurs in four phases [15, 31]. The first phase, which is termed scientific convergence, occurs when distinct scientific fields begin citing each other, leading to an increase in cross-scientific research. It is followed by the technology convergence phase in which previously distinct technological fields increasingly overlap and new technology platforms arise. As technology convergence continues to blur existing technological and market boundaries, market convergence, the third phase, takes place, resulting in the emergence of new product-market combinations. In some cases, the convergence of technologies and markets may impact existing industries such that a fourth phase in form of the convergence of entire industries occurs. A typical and relatively recent example of convergence is the emergence of the smartphone, which nowadays combines initially separate technologies such as phones, cameras, and portable computing devices. It eventually created an entirely new market, completely transformed the mobile phone industry, and even overthrew the compact camera industry. Another example is Bitcoin [7], which was created by using techniques from various computer science domains such as distributed systems, cryptography, security, and game theory.

Concerning a potential convergence of DLT and AI, we currently see the emergence of the first scientific publications that do not simply apply AI in the context of DLT (e.g., deep learning for prediction of the Bitcoin price) or vice versa, but instead discuss a deeper integration of both technologies. Extant literature that provides a consolidated overview of the integration possibilities of AI and DLT, however, is scarce. Dinh and Thai [13] provide a viewpoint with conceptual ideas of how an integration of AI and blockchain technology might look like. They outline possibilities for the reciprocal integration of AI and DLT, highlighting that AI can be used for blockchain, just as blockchain can be used for AI. Specifically, AI can support blockchain systems by increasing their security and scalability, by acting as an automated referee and governance mechanism, and for privacy-preserving personalized systems. Blockchain, on the other hand, can serve AI systems by enabling decentralized computing, providing data sharing infrastructures, serving as a trail for explainable AI, or by coordinating devices in the Internet of Things. As the article is a viewpoint, the authors remain on a rather conceptual level and do not provide an in-depth review of extant literature. Salah, et al. [12], on the other hand, provide a review and open research challenges on one of these two perspectives, the use of blockchain for AI. The authors identify enhanced data security, improved trust on robotic decisions, business process efficiency increases, collective decision making and decentralized intelligence as main drivers for the usage of blockchain for AI applications. Interestingly, the latter category has similarities with the application of privacy-preserving personalization, which Dinh and Thai [13] classify as an AI for blockchain perspective. Furthermore, Salah, et al. [12] develop a taxonomy that provides an overview of different relevant technical characteristics, such as the consensus protocol or the blockchain type. Their review surveys research articles, as well as industry whitepapers. As such, it delivers valuable insights into how blockchain can be used for AI in vertical industry use case cases, such as banking, finance, or agriculture. The article concludes with several open research questions on technologies that are relevant in the context of blockchain for AI applications. These include topics of increasing blockchain systems' scalability and data confidentiality, interoperability, or quantum cryptography. Lastly, Karafiloski and Mishev [32] provide an overview of how blockchain can be used for storing and organizing large amounts of data. Although the authors do not consider AI use cases in much detail, the presented work is representative for some fields of the convergence of AI and DLT, such as data sharing.

In summary, prior research lacks an understanding how AI and DLT are reciprocally integrated in systems today, and how future research can advance the convergence of these technologies. Specifically, prior research focuses on blockchain, a specific DLT concept, and not on potentially other DLT concepts that may be better suited in the context of AI [12, 13]. Furthermore, prior research only provides purely conceptual ideas how AI may be used for DLT, and does not evaluate today's technical feasibility of such systems [13].

## III. METHODOLOGY

### A. DATA COLLECTION

For the identification of articles addressing the convergence of AI and DLT, we systematically searched scientific databases. To cover a wide range of journal and conference publications, we queried IEEE Xplore, ACM Digital Library, AIS Electronic Library, Science Direct, Scopus, and Ebsco Business Source. Our search string required the publications to have a DLT-specific term and an AI-specific term in either their title, abstract, or keywords. The search string we employed was:

TIKEAB ((Blockchain OR „Distributed Ledger" OR DLT OR Bitcoin OR Ethereum OR Cryptocurrency OR „Crypto currency" OR „Crypto-currency" OR „block chain" OR „Smart contract") AND (AI OR ML OR „Artificial Intelligence" OR „Machine Learning" OR „Deep Learning" OR Clustering OR Classification OR „Neural Network" OR „Big data" OR „Data mining" OR „Intelligent system*" OR „Statistical model", OR „Statistic model")).

We excluded articles published before 2008, since the concept of Bitcoin emerged that year. For DLT, the search string included its most common concept of blockchain, as well as blockchain's most frequent implementations (i.e., Bitcoin and Ethereum) and other technical terms. For AI, the search string included ML and the most common application forms of it. We searched on December 19th, 2019, this resulted in 2,411 unique articles.

In a first step, we screened each of the 2,411 articles title, abstract, and keywords to check whether they actually covered AI and DLT, and, whether they were written in English. As not all of the databases offer the option to exclude articles published before a certain year, we also manually checked whether the articles were published in or after 2008. For articles, whose title, abstract, and keywords provided inconclusive information only, we also checked their full text in a second screening step. Overall, this procedure led to the exclusion of 2,224 articles that did not pass the filter criteria. After that, we analyzed whether the remaining 187 articles described a close integration of AI and DLT according to the concept of convergence [14, 15] and whether these articles answer the purpose of using AI and DLT. During this third step, 32 articles eventually turned out to be relevant for further analysis. A large share of the articles which did not consider the close integration of AI and DLT, according to the lens of convergence, covered AI-based cryptocurrency price forecasting or AI-based trading bots for cryptocurrency exchanges.

### B. DATA ANALYSIS

Following the data collection, which resulted in the identification of 32 relevant articles, we categorized all 32 articles into groups. The groups were thereby derived from Dinh and Thai [13]'s viewpoint and adapted to our review, where necessary. Toward this end, we expanded the concept of blockchain to DLT, and reframed the focus and name of some categories to better suit the extant literature (e.g. *secure DLTs* instead of *secure and scalable DLTs*, and *coordination of devices* instead of *coordination of untrusting devices*). Table 1, below, provides an overview of the adapted coding scheme and the number of articles in each category. Note that we categorized some articles into multiple subcategories. Therefore, the sum of articles in Table 1 is higher than 32. We also added another type of perspective, *Both*, to the coding scheme. It consists of consolidating works, such as literature reviews or high-level articles covering both aspects of the convergence of AI and DLT. Our future research agenda is an extension from our review and draws from multiple sources. On the one hand, the findings, outlook and conclusion of the extant articles discussed in our review. On the other hand, our own assessment of the recent developments in the fields of AI and DLT.

TABLE 1. Classification of the identified articles that cover the integration of AI and DLT into groups.

| Perspective | Group | Number of articles |
|---|---|---|
| AI for DLT | Secure DLTs | 7 |
| | Automated referee and governance | 3 |
| | Privacy-preserving personalization | 6 |
| DLT for AI | Decentralized computing for AI | 13 |
| | Secure data sharing and marketplace for AI | 14 |
| | Explainable AI | 6 |
| | Coordinating devices | 1 |
| Both | Other | 2 |

## V. REVIEW ON AI FOR DLT

Drawing on the general distinction between AI for DLT and DLT for AI proposed by Dinh and Thai [13], this section describes our findings in terms of how extant research has applied AI for DLT. We identified three different groups of use contexts, which we detail below, and summarize our findings in Table 2.

### A. SECURE DLTS

1) DLT PROTOCOL SECURITY

Within this category, extant literature solely applies AI methods of reinforcement learning to explore and develop strategies in game theoretic settings. Specifically, these articles analyze the fairness of mining activities. The learning agents get rewarded with a cryptocurrency-based miner reward. Within the category of articles using reinforcement learning for DLT protocol security, two subcategories exist.

*Selfish mining.* Articles within this subcategory analyze selfish mining strategies for blockchain systems [10, 33]. The learning agent thereby is a blockchain miner, which can strategically delay the publication of found blocks. By doing so, the selfish miner aims to waste mining power of honest miners that work on another fork of the blockchain [45]. Through generating these attacks on blockchain systems in a testing environment using reinforcement learning, researchers can find new insights on the security of these blockchain protocols. For example, Hou, et al. [10] performed the simulation with protocols such as Bitcoin [7], Ethereum [29], or GHOST [46]. In their research, an agent learns mining strategies where it can gain disproportionately large mining rewards relative to its hash rate. While some of these adversarial mining strategies are known from theoretical studies of the blockchain protocols [45], AI has the potential to detect new, previously unknown attacks in complex scenarios, for example, in cases with multiple partially cooperating agents. Additionally, the simulation may contribute to a better understanding of the feasibility of these mining strategies [10].

*Pool operation.* In research within this subcategory, the agent learns strategies for operating a mining pool. In our review, we identified only one article that fits this subcategory: Haghighat and Shajari [34] specifically analyze a block withholding game, where a pool operator can either decide to mine honestly, or attack another pool by mining (at least with a fraction of its power) for the other pool, but not actually submitting a solution, in case a block is found. As a consequence, the revenue of the attacked pool drops, and it gets less attractive for other, honest miners to mine as a member of this attacked pool. The authors train an agent, which represents a pool operator, with reinforcement learning methods. It decides between mining honestly and attacking other pools at each step in the game. One insight from the results is that it may be more likely than widely assumed that one pool operator at a certain point in time was in control of more than 51 % of Bitcoin's network mining power [34].

TABLE 2. Classification of the identified articles that cover the integration of AI and DLT into groups.

| Group and categories | Summary of concepts | Exemplary sources |
|---|---|---|
| Secure DLTs | | |
| DLT protocols | • Study of fairness of game-theoretic mining scenarios with reinforcement learning and implications on blockchain security<br>• Some articles focus on selfish mining strategies<br>• Other articles focus on mining pool operation | • [10, 33, 34]<br>• [10, 33]<br>• [34] |
| Smart contracts | • Passive smart contract analysis using bytecode or metadata analysis<br>• Active smart contract invalidation using reinforcement learning | • [35-37]<br>• [38] |
| Automated referee and governance | | |
| DLT protocols | • Automatically govern DLT protocol parameters or transactions | • [39, 40] |
| Smart contracts | • Assist users in votes of changing SCs | • [41] |
| Privacy-preserving personalization | | |
| Local computation | • DLT as an auditable data sharing infrastructure, computing ML model inference locally, federated learning for model updates | • [42, 43] |
| Trusted Execution Environments | • DLT as an auditable data sharing infrastructure, TEEs for confidential and computationally intensive SCs, ML training and inference in the SCs<br>• Federated learning for increased data security<br>• Differential privacy for increased data security | • [11, 44]<br>• [44]<br>• [11] |

2) SMART CONTRACT SECURITY

Articles within this category aim to protect users from interacting with insecure or malicious smart contracts. Two subcategories with different approaches exist in the extant literature: First, the analysis and detection of such smart contracts. Second, the active interaction with such smart contracts in order to manipulate and invalidate them.

*Detection.* Three articles in this subcategory aim to detect security issues in smart contracts using data analysis techniques. As the smart contract opcode is usually published on the distributed ledger, two articles [36, 37] analyze the opcode with neural networks. Since a compiled opcode is challenging for humans to read and understand, Kim, et al. [36] use neural networks to estimate the functionality of a given smart contract (e.g., whether it is intended for a marketplace or for gaming). Tann, et al. [37], in contrast, aim at detecting security vulnerabilities in smart contracts. To do so, they run long short-term memory neural networks on a sequence of opcodes. In a performance comparison with a symbolic analysis approach, the machine-learning-based analysis method can outperform the formal verification-based method with regards to classification accuracy and speed. Another article by Camino, et al. [35] aims at detecting honeypot smart contracts. These types of smart contracts appear to contain free, withdrawable funds. However, once a user aims to invoke the honeypot contract to redeem free tokens, the honeypot actually does not release the funds, resulting in the user losing funds that they may have used to invoke the honeypot contract. Contrary to the other articles in this subcategory, Camino, et al. [35] do not use opcode analysis techniques, but analyze the smart contract metadata (i.e., values derived from related transactions and fund flows) and off-chain data (i.e., the availability of a source code and the number of source code lines). In this way, data science techniques can classify more than 85 % of unseen smart contracts correctly. This approach especially works well in cases where there was a minimum of smart contract on-chain activity, and, as a result, on-chain metadata is available.

***Manipulation.*** The only article identified within this subcategory [38] uses an active reinforcement learning approach to invalidate criminal smart contracts. Criminal smart contracts can be used for illegal activities, such as selling confidential information [47]. In this research, the agent learns to manipulate the contracts' data feed and thereby successfully invalidates a substantial share of the studied criminal smart contracts. While doing so, the system aims at invalidating these given smart contracts, but not at detecting possible criminal smart contracts in the first place.

## B. AUTOMATED REFEREE AND GOVERNANCE

Another, currently small group of articles uses AI for DLT-based system governance. Dinh and Thai [13] present the vision that people, devices, and smart contracts record transactions on a distributed ledger. An AI can then solve potential disputes of events happening on- or off-chain, and record the results on a distributed ledger. This automated arbitration could be data driven, unbiased, and, as a result, more consistent, justified, and accepted than arbitrations today [48]. In extant literature, articles toward this vision appear to be in an early stage and fall within two categories: AI-based DLT protocol governance, and AI-based smart contract governance.

### 1) DLT PROTOCOL GOVERNANCE

This category currently consists of two articles. Lundbæk, et al. [39] propose an adjusted proof-of-work-based consensus mechanism. Only a subset of nodes participates in the required hash calculation and the DLT protocol uses ML to regularly update system governance parameters, such as the ideal number of miners or the level of mining difficulty. In this scenario, AI is tightly integrated with the DLT protocol. However, the authors do not discuss in detail the functionality and security aspects of their ML-based governance. Gladden [40], on the other hand, discusses ethical aspects about cryptocurrencies governed by an AI system. The author claims that such a system can positively influence the ethos and values of societies. However, the article has a sociotechnical focus and does not describe the technical architecture of the system.

### 2) SMART CONTRACT GOVERNANCE

Only one article in the extant literature fits this category. Liu, et al. [41] propose to implement voting mechanisms for participants in smart contracts to alter smart contract parameters. As such, these voting mechanisms can govern smart contracts in complex situations and can alter the smart contracts towards an adaptive behavior. An ML-based system assists the users in voting, based on their past voting behavior. As a result, the ML-based system pre-chooses the users' selections and eases their tasks. In this scenario, the ML-based system runs outside of the actual DLT system.

## C. PRIVACY-PRESERVING PERSONALIZATION

Many internet platforms nowadays collect data about their users and apply AI-based recommender systems to personalize their content. Examples include Facebook, Netflix, or Taobao in China. A recommendation by this AI-based system thereby is not only influenced by the individual user's data, but also by all the other users' data. Netflix, for example, can recommend new movies to users based on their individual and other users' watching behavior. However, this comes at the risk of impeding users' privacy. In several cases, private data from such platforms has been leaked to the public [49] or misused [50]. A group of articles envisions AI-based personalization for DLT-based data sharing platforms. This could, for example, include a social network built on a DLT infrastructure [13]. DLT can thereby serve as a transparent trail of data flows, and give users control over their data. In the extant literature, two categories of articles exist with different approaches toward designing such systems. First, a category with articles that aim to use local computation. Second, a category with articles that use distributed ledgers based on hardware-assisted Trusted Execution Environments (TEEs).

### 1) LOCAL COMPUTATION

An ML model inference intended to personalize content on a platform requires data about the user to personalize its recommendation for them. If such a model inference is executed locally, there is no need for the user to share their data, while they can still get a personalized recommendation. However, if no user shares their data with the platform, it is challenging for a platform operator to train ML models using traditional methods. Therefore, articles in this category describe systems that use federated learning. With this distributed ML technique, an ML model gets trained locally on a user's device and only ML model updates leave the device and are shared with the platform or potentially other parties. Model updates from many users are then aggregated into a new model. Federated learning is already successfully applied on a large scale on smartphones, for example, to predict the next word a user may want to type on their keyboard [51]. The distributed ledger provides an infrastructure to share data on an auditable and immutable ledger. Some articles describe systems that only store hashes of model gradient updates or hashes of aggregated models on the distributed ledger, in order to save storage space on the ledger and preserve data confidentiality [52]. Other articles describe systems that store the complete gradient updates and aggregated models themselves on the ledger [53]. First applications are illustrated for the Internet of Things in general [42] or the taxi industry [43].

2) TRUSTED EXECUTION ENVIRONMENTS

The second category uses TEE-based DLTs. TEEs are typically located on a CPU and provide a special enclave for secure computation. In this enclave, applications are executed such that other applications executed on the same CPU but outside the enclave cannot interfere with the state or control flow of the application shielded by the TEE. As a result, the enclave appears as a black box to the outside and secures the application running inside. Furthermore, the hardware of the TEE can generate a proof to the application running inside the TEE to attest that the software is executing on this specific, trusted hardware. This feature is called remote attestation. Extant research uses DLT protocols to coordinate TEEs, thus enabling confidential smart contracts that execute inside the TEE. Since the TEE ensures data confidentiality, this category does not necessarily require local computation. First articles aim at providing solutions for the medical industry [11, 44]. To further increase data security, some articles describe systems that combine TEEs with federated learning [44] or differential privacy [11]. Simplified, differential privacy is a mechanism that adds randomly distributed noise to a set of data points. This protects the information privacy of users that provide data. As the random noise has zero mean (or a predefined mean), an aggregator can still draw meaningful conclusions from the aggregated data set without threatening anyone's information privacy.

## V. FUTURE RESEARCH AGENDA ON AI FOR DLT

In this section, we present our analysis of open, future research opportunities on the advancement of DLT using AI. Again, we use the categories proposed by Dinh and Thai [13] to structure the future research agenda. We summarize our findings in Table 3.

### A. SECURE DLTS

Dinh and Thai [13] have drawn a futuristic picture with far-reaching real-time analysis and decision possibilities of an AI as part of the DLT system. Current AI-based systems, however, do not provide the required security and robustness guarantees necessary to govern a DLT system. Precisely, while AI-based systems can detect software vulnerabilities, they cannot (yet) guarantee that it has identified all available security vulnerabilities. If an AI detects no vulnerability, it does not necessarily mean that there are none existing [10]. This open research problem of AI robustness is also highly relevant in other AI application domains, such as autonomous driving, where first results may be transferable to software and DLT systems security [56]. In general, methods from the field of explainable AI (XAI) [19] could help to better understand how reliable an AI-based decision is. We, therefore, see this as a fruitful field for future, foundational research. While AI is a promising technology to detect DLT security vulnerabilities, further research that aims to resolve these vulnerabilities and to build secure DLT-based systems is required in parallel [34]. Subsequently, we present future research opportunities in the two categories identified in our review in chapter four. Furthermore, we see possible research opportunities in a convergence of these two categories, which we also present.

1) DLT PROTOCOL SECURITY

The security analysis of DLT protocols using AI is currently dominated by reinforcement learning. Extant literature analyzes game-theoretic incentive mechanisms in widely adopted protocols such as Bitcoin or Ethereum. Possible future research opportunities include the expansion of previous work with settings such as partially cooperating miners [10], or the analysis of strategies other than selfish mining [33]. In our view, beyond analyzing the fairness of mining settings and its DLT security implications, another additional interesting field of AI-based security analysis is the DLT source code itself. This is a highly relevant field in practice. For example, the deployed and widely-used cryptocurrency Zcash, which uses a novel form of zero knowledge proof cryptography, has been subject to a vulnerability that would have allowed an attacker to generate an infinite amount of cryptocurrency tokens [57], thus potentially inflating the assets of other users. Extant literature already uses AI-based analysis methods for software bug identification in general [54], and smart contract opcode analysis in particular [36, 37], thereby often outperforming formal verification-based methods [37]. We, therefore, see a promising future research avenue in further developing and applying such AI methods for analyzing the DLT protocol code security.

TABLE 3. Overview of future research opportunities in the field of AI for DLT.

| Group and categories | Future research opportunities | Exemplary sources |
|---|---|---|
| Secure DLTs | | |
| DLT protocols | • Advance reinforcement learning research for game-theoretic analysis<br>• Use supervised learning methods for protocol software quality evaluation | • [10, 33, 34]<br>• [36, 37, 54] |
| Smart contracts | • Assist developers in developing secure SCs<br>• Apply reinforcement learning for game-theoretic analysis of complex smart contracts<br>• Refine supervised learning methods for SC security analysis (e.g., with regards to robustness) | • [35-38]<br>• [38]<br>• [35-37] |
| Interaction of DLT protocols and smart contracts | • Explore game-theoretic interplay between complex smart contracts and DLT protocols by means of reinforcement learning | • [10, 55] |
| Automated referee and governance | | |
| DLT protocols | • Advance AI robustness, evaluate system security | • [39] |
| Smart contracts | • Advance AI robustness, extend systems and evaluate security | • [41] |
| Privacy-preserving personalization | | |
| - | • Development of data and ML model sharing platforms<br>• Application of practical data privacy technologies, such as TEEs combined with federated learning and differential privacy<br>• Deployment and evaluation within trusting consortia | • [11, 44]<br>• [11, 44]<br>• [44] |

2) SMART CONTRACT SECURITY

Extant literature in this subcategory mostly aims to protect users from insecure [36, 37] or malicious [35, 38] smart contracts. Practical experience, however, has shown that developers struggle to develop secure smart contracts in the first place [58]. Therefore, an interesting perspective for future research would be in the field of using AI to assist developers in developing secure smart contracts. This could include the development of early security warning tools for developers that automatically check developed code for security vulnerabilities. Recent work in other fields of software engineering suggests that such AI-based systems may be feasible [54, 59].

Beyond this suggestion to expand the usage perspective of future research, we see further research opportunities in refining and applying the methods identified in our review. Reinforcement learning, which has already shown promising results for DLT protocol security analysis in simulation settings [10, 33, 34], appears as a promising method to analyze the security of game-theoretically complex smart contracts [38] in simulations. This could include, for example, decentralized token exchanges.

The other identified approach is to analyze security vulnerabilities in smart contracts by supervised learning. This appears as a promising field of AI application, as AI-based methods have outperformed classical, formal verification-based methods [37]. Future research could expand this work by considering further classes of smart contracts [35] and further details, such as different smart contract compiler versions [36]. Furthermore, future research that aims to support software developers could aim at not only detecting security vulnerabilities in a smart contract, but also at localizing this vulnerability (e.g., by expressing the information which portions of the bytecode cause the vulnerabilities [37]) or even by providing suggestions to fix it [60, 61].

3) INTERACTION OF DLT PROTOCOLS WITH SMART CONTRACTS AND SECURITY IMPLICATIONS

In addition, we see a convergence of these two categories—AI for DLT protocol security, and AI for smart contract security—as a promising avenue for future research. Recent research has investigated the extent to which game theoretical aspects in DLT protocols and smart contracts influence each other and can lead to unfair conditions for regular users of the overall system [55]. For example, miners who may participate in the trading of assets on blockchain-based decentralized exchanges have the sovereignty to decide which transaction is included in a block and which one is not. This can incentivize other users to pay higher transaction costs for these exchange transactions than they would without the miner participation. Prior research considers this as an unfair setting [55]. This is just one example of the complex interplay of smart contracts and DLT protocol incentives, which is also relevant in decentralized applications other than token exchanges [62, 63]. As it appears to be an even more complex topic than the standalone security of DLT protocols or smart contracts, an AI-based system could provide valuable insights. Prior research uses reinforcement learning for game-theoretic analyses of DLT protocols and smart contracts separately, therefore, reinforcement learning also appears as a promising method to analyze the interplay between both. The results could be used to support the development of fair and secure DLT-based systems.

## B. AUTOMATED REFEREE AND GOVERNANCE

In general, both identified categories—AI for smart contract and for DLT protocol governance—face the reality of current AI's capabilities. To automatically perform far-reaching decisions on a distributed ledger with a potential influence on financial or data confidentiality circumstances, most of today's AI's robustness and explainability guarantees are not strong enough. The robustness guarantees of many modern AI systems are weak [64] compared to the complexity of the actions that agents perform when interacting with DLT-based systems. Therefore, in our view, breakthroughs in the robustness, explainability, and ultimately security of AI systems are required before they can automatically govern DLT protocols and smart contracts on a large scale. However, in the case of human-in-the-loop based AI smart contract governance [41], an AI may assist a human decision maker which can intervene at any time in case an AI system faces and detects an irregular situation. This human-in-the-loop model, therefore, appears to provide good practicality with today's AI technology. In our view, extant research provides a good starting point for future research, especially for AI-based human-in-the-loop smart contract governance [41].

## C. PRIVACY-PRESERVING PERSONALIZATION

DLT-based platforms for data sharing are starting to get more attention in research but are not deployed and scaled in practice yet. As such, research on such platforms and their AI-based personalization is still in an early stage. Articles in the subsection of privacy-preserving personalization using AI for DLT often incorporate aspects from multiple other subsections, such as *decentralized computing for AI*, or *secure data sharing and marketplace for AI*. In our view, developments in the field of *privacy-preserving personalization* depend on advancements in the underlying secure computation and privacy-preserving data sharing technologies.

An interesting future research opportunity which is practical with today's technologies is the further development and deployment of DLT-based data sharing platforms and an AI deployment on top of these platforms. TEEs appear as a promising technology to realistically build such systems today, due to their relatively small computational overhead when compared with other secure computation technologies, as well as their ability to enforce policies [65]. TEEs can be combined with federated learning for stronger data confidentiality [44]. A promising avenue for deployment appears to be within consortia of (at least partially) trusting participants. This could, for example, include a consortium of hospitals aiming to deploy personalized, AI-based treatments while complying with strong privacy requirements [44].

## VI. REVIEW ON DLT FOR AI

Drawing, again, on the general distinction between AI for DLT and DLT for AI proposed by Dinh and Thai [13], this section describes our findings in terms of how extant research has applied DLT for AI. We identified four different groups of use contexts and summarize our findings in Table 4.

TABLE 4. Overview of the concepts in the extant literature using DLT for AI.

| Group and categories | Summary of concepts | Exemplary sources |
|---|---|---|
| Decentralized computing for AI | | |
| DLT protocol | • Mechanism for block candidate selection, training AI models instead of calculating hashes, separate group of task providers, reputation mechanisms to secure the system | • [26, 66, 67] |
| Smart contracts | • Scaling SC's capabilities (e.g., through TEEs), training AI models in these smart contracts while ensuring the computational integrity | • [11, 44, 68, 69] |
| Off chain | • Training AI models off-chain through federated learning<br>• Partially, application of game-theoretic incentive mechanisms to ensure computational integrity | • [42, 43, 53, 70]<br>• [71] |
| Secure data sharing and marketplace for AI | | |
| Smart contracts | • Manage access to data on decentralized file protocols through DLT<br>• Financial rewards and staking mechanisms to incentivize honest behavior | • [72]<br>• [73] #235} |
| TEEs | • Execute computationally intensive SCs in a TEE<br>• Manage a large share of the ML pipeline through the system | • [11, 44, 69]<br>• [11, 69] |
| Federated learning | • Store model update plain data or hashes on the ledger, increase federated learning transparency and accountability | • [42, 43, 53, 70] |
| Explainable AI | | |
| - | • DLT as a trail for model training and inference (e.g., for federated learning)<br>• In some articles, TEEs to enforce policies | • [11, 44, 52, 53, 69]<br>• [11, 44, 69] |
| Coordination of devices | | |
| - | • Distributed ledger for device coordination (e.g., through public keys and reputation management)<br>• Remote attestation of a TEE | • [74]<br>• [75] |

### A. DECENTRALIZED COMPUTING FOR AI

Due to the large amount of processed data, ML model training is often computationally expensive. Nowadays, graphics processing units (GPUs) are typically used to train ML models, as they provide more computational power for most ML training tasks. For complex optimization tasks, many GPUs may be necessary to train ML models in a reasonable amount of time [76]. However, many GPUs and CPUs in computers around the world are only slightly loaded or even unused. Previously, distributed computing has been applied to utilize these unused resources in communities at scale, for example, to perform protein folding calculations for medical research [77]. Articles within this group describe systems that use decentralized computation methods to train AI models. DLT can serve as a trail to organize this decentralized computation, as well as provide a ledger for rewards paid in cryptocurrency [13].

Looking at the extant literature, three different approaches for DLT-enabled decentralized computing for AI exist: First, the computation within the DLT protocol. Second, the computation in smart contracts. Third, the computation outside the distributed ledger.

1) DLT PROTOCOL

Some deployed blockchain networks, such as the public Bitcoin or Ethereum main network, are secured by a proof-of-work-based consensus mechanism. This describes the search for a number such that the hash of a blockchain block candidate, which includes that number, is below a certain threshold. This mechanism is required to limit the number of nodes that can propose a new block, and ultimately, to ensure blockchain security. By adjusting the threshold, the network can adjust the difficulty of this search problem, and ultimately adjust the average time to find a new block. The search problem is computationally expensive, however, the verification of a potential solution is computationally inexpensive.

The proof-of-work mechanism consumes large amounts of energy for protocols such as Bitcoin, which some authors see critical [66]. At the same time, the calculated hashes cannot be meaningfully used for purposes other than securing the network. Some authors, therefore, propose a proof-of-useful-work mechanism, in which miners train an ML model instead of finding a number for a certain block hash. The block which contains an ML model with the least test error is then accepted as the new block by the other nodes [67]. For such a system to work, however, several challenges have to be overcome. On the one hand, the computational difficulty of optimizing an ML model, is hard to adjust [78]. This can also cause blockchain forks to occur on a very regular basis. On the other hand, the proof-of-useful-work mechanism for AI requires the secure provision of an ML model architecture, training data, and test data. Extant literature that describes such systems uses selected groups of participants and committees with reputation mechanisms to govern this data provision [26, 67, 78].

2) COMPUTATION IN SMART CONTRACTS

In some DLT networks, such as Ethereum, a virtual machine can execute Turing-complete smart contracts. As such, DLT could provide a substrate for ML model training [68]. However, traditional ledgers such as Ethereum do not support intensive computations that would be necessary to perform ML model training [71]. To overcome this limitation, some articles propose to extend distributed ledger smart contract execution with TEEs such that these ledgers can support computationally more intensive smart contracts. This TEE can then train simple ML models [11, 69].

3) OFF-CHAIN COMPUTATION

Another subcategory of articles performs computations not on the distributed ledger, but off-chain. Unless using sophisticated cryptographic techniques [79], the integrity of the computation cannot be guaranteed. DLT-based federated learning systems, for example, train an ML model on the edge [42, 43, 53, 70]. The main motivation of most articles focusing on DLT-based federated learning, however, is not the usage of decentralized edge computing resources per se. Instead, the main motivation is secure data sharing. Some of the DLT-based federated learning articles also propose financial rewards for both in combination, model training and data sharing [80, 81]. Besides federated learning, another article by Lu, et al. [71] proposes a crowdsourced computation approach to offload heavy computation tasks from a blockchain, such as in ML model training. Multiple offloaded computing engines audit each other's work and game theoretic incentive mechanisms are used to build a protocol. The authors also present a security analysis of their protocol.

*B. SECURE DATA SHARING AND MARKETPLACE FOR AI*

In addition to the increasingly available computing power, another fundamental reason for the recent advancement in AI is the strong growth of available and digitized data. ML-based systems generally perform better the more data they are trained on, for example, with regards to classification accuracy [21, 82]. Some authors see recent developments in IT, in which few companies are in control of large amounts of personal data, critical and propose DLT-based data markets to democratize such data silos [13, 83]. As such, a group of articles proposes solutions based on DLT to build data sharing infrastructures, thus enabling the deployment of AI. These articles differ in their technical approaches toward designing such systems. Articles which incorporate advanced privacy-preserving mechanisms in their systems mostly focus on the health care industry. The reason for this are potentially the strong confidentiality requirements on health data.

### 1) SMART CONTRACTS

Two extant articles use traditional smart contracts on Ethereum to build a data sharing infrastructure and marketplace. These articles differ in their focus, either on sharing training data itself [72], or ML models [73]. Especially the sharing of training data requires the storage and handling of large files. Therefore, Özyilmaz, et al. [72] connect their system to the decentralized file protocol SWARM and manage data access rights through the blockchain. These marketplaces allow the payment of data providers with units of the cryptocurrency Ether. To incentivize participation with high quality data, Harris and Waggoner [73] propose staking mechanisms in which malicious participants sharing spam models lose their stake. Such systems can be applied in the Internet of Things industry in general [72].

### 2) TRUSTED EXECUTION ENVIRONMENTS

TEEs enable the computation of relatively intensive tasks, while preserving data confidentiality and integrity throughout the computation. Therefore, systems described in some articles execute computationally intensive smart contracts off-chain in a TEE [11, 44, 69, 84, 85]. These articles use the blockchain concept with different DLT designs [3]. Several articles use the Oasis blockchain [11, 84, 85]. Hynes, et al. [11], for example, present a privacy-preserving data market that provides solutions for a large share of the ML pipeline. Through smart contracts, data providers can define policies to share their data. These policies, for example, include the asking for a reward and differential privacy requirements. Data consumers can choose to fulfill these policies in order to train an ML model on the providers' data. Since a TEE ensures confidential computation, the training data does not get leaked and only data consumers can get access to ML model inference. The ML model itself is shielded inside a smart contract and inference executions count toward the provider's policies, which increases data provider's privacy against potential inference attacks. This type of attacks aims at executing the ML model in order to extract the underlying training data or the model itself [86].

Other articles, which describe a system with similar capabilities, use an Ethereum-based blockchain [44, 69]. As a special feature, a virtual machine in the enclave can allow the training of proprietary AI models while not comprising the training data security [69]. Passerat-Palmbach, et al. [44] incorporate federated learning into the TEE-enabled blockchain. This protects data providers from potential TEE side channel attacks and, therefore, further reduces data privacy risks.

### 3) FEDERATED LEARNING

A large subcategory of articles describes systems that use federated learning with distributed ledgers that are not TEE-enabled [42, 43, 52, 53, 81]. DLT thereby serves as a provenance record of data. This data can describe a large share of the ML pipeline —training data origin, training data, ML model modifications, or testing data [52]. In most systems described in these articles, the data itself is not stored on the blockchain, but hashes of the data [52]. In some cases, the systems use relatively simple ML models and store plain model updates on the blockchain [53]. As a result, every participating server can audit and compute the aggregated ML model weight updates. A blockchain-based solution can be well-suited for certain use cases and only provides a small performance overhead of ca. 5 % to 15 %, while enabling transparency and accountability [53]. Some systems go further and replace the traditionally centralized aggregator with a smart contract-based one [43].

As federated learning is potentially vulnerable to inference attacks [86], some articles describe systems that further use differential privacy techniques [42, 70] to increase data confidentiality guarantees. Further systems incorporate financial incentives for participants sharing model updates [42, 81].

### C. EXPLAINABLE AI

Complex ML models, such as deep neural networks, are nowadays often used in a black box manner [19]. This means that users or even system creators do not have the information how these models come to a certain prediction. However, obtaining this information can be desirable in certain cases, for example, to verify the system's robustness or to comply with legislation [19]. Dinh and Thai [13] outline DLT as a technology to increase AI systems' explainability. Their vision is that DLT provides an immutable trail to track the data flow of AI-based systems.

Looking into the extant literature on DLT for AI, the explainability of the black-box model itself is out of focus for most researchers. The DLT for XAI literature mainly covers data provenance or computational integrity aspects for model training or inference. Sarpatwar, et al. [52] design a DLT-based federated learning system for trusted AI and present five requirements of blockchain for trusted AI. First, guarantees of AI model ownership and track of use are important. Second, the confidential sharing of AI assets as they are often created using sensitive data. Third, auditability regarding the AI training process. Fourth, the traceability of the AI pipeline. Fifth, to keep a record in order to recognize a potential bias and model fairness issues.

In general, several extant articles aim to use DLT-based federated learning to ensure such aspects of AI model explainability and trustworthiness [52, 53, 73]. Other articles use TEEs to also cover the model inference with their system, and thus, provide stronger explainability and restrictions in order to track data flows for data providers [11, 44, 69]. Hynes, et al. [11], for example, use XAI methods to compensate data providers relative to their influence their data had on an ML model's inference.

### D. COORDINATION OF DEVICES

A final group of articles aims to coordinate devices with DLT. These devices, such as Internet of Things devices, generate data, which an AI system can potentially analyze. A distributed ledger can serve as a registrar for these devices, serve as a storage for metadata (such as hashes), or even as a data storage. In extant literature, articles that use DLT also connect devices to the distributed ledger. Using cryptographic principles, transactions on a distributed ledger (such as smart contracts) are signed, which ensures that only legitimate participants can perform certain transactions. Kang, et al. [74] propose to use DLT for reputation management in a federated learning setting. Beyond identification through asymmetric cryptographic mechanisms on a distributed ledger, TEEs provide mechanisms to self-identify and remote-attest [75], such as physically unclonable functions [87].

### VII. FUTURE RESEARCH AGENDA ON DLT FOR AI

In this section, we present our analysis on open, future research fields on DLT for AI. Again, we use the categories based on Dinh and Thai [13] to structure the future research agenda. We present an overview of our results in Table 5.

### A. DECENTRALIZED COMPUTING FOR AI

As we identified three categories of approaches using DLT for decentralized computing for AI, the future research opportunities in these approaches differ. Subsequently, we present our identified research opportunities in all of these.

#### 1) DLT PROTOCOL

First extant research presents blockchain designs with proof-of-useful-work mechanisms that could be used for ML model training. Proof-of-useful-work itself has a long history with several blockchain-based cryptocurrency systems deployed [98], for example, for number theoretic research [99]. While this prior work is interesting from an academic point of view, the deployed cryptocurrencies so far had little practical relevance based on their market capitalization. Some blockchain designs, such as the Ethereum mainnet, actually aim to not use consensus mechanisms based on proof-of-work in the future [100].

Despite that, we see promising avenues for future research on proof-of-useful-work for AI, as well as for actually deploying and evaluating such systems. On the one hand, future research could analyze the economics of such a system and its utility for ML model requesters as well as regular DLT users. On the other hand, future research could analyze the practical security guarantees of such a system. Aspects include the security of incentive mechanisms, as well as the security of DLT transactions. This is particularly interesting, since an attacker with large computational power would not only benefit from rearranging transactions on the ledger, but also potentially from the useful work itself.

#### 2) COMPUTATION IN SMART CONTRACTS

Extant research has aimed to compute smart contracts off-chain in TEEs, which enables a computation with maintained integrity. Out of several approaches to off-chain computation while ensuring integrity [79], TEEs are interesting due to the relatively high computational power they provide [88].

TEEs have seen prior research from both fields separately, AI and DLT. In the field of AI, extant research provides frameworks for ML model training and inference on TEEs [90]. It has shown that TEEs are in general capable of simple practical machine learning tasks, such as speech processing [101], or the processing of small images [88]. In the field of DLT, on the other hand, researchers aim to increase the low performance of smart contracts using TEEs while ensuring confidentiality [75].

TABLE 5. Overview of future research opportunities in the field of DLT for AI.

| Group and categories | Future research opportunities | Exemplary sources |
|---|---|---|
| Decentralized computing for AI | | |
| DLT protocols | • Further develop, deploy, and evaluate systems, analyze incentive mechanisms, analyze system security | • [66, 67] |
| Smart contracts | • Develop systems to enable scalable SCs through TEEs<br>• Improve TEE performance and security<br>• Evaluate suitability of systems for practical use cases<br>• Incorporate privacy mechanisms into computing | • [44, 69, 88]<br>• [89]<br>• [11, 44, 88, 90, 91]<br>• [95] |
| Off-chain computation | • Deploy DLT-based, federated learning especially within trusted consortia<br>• Further research on game-theoretic incentives for secure computation<br>• Foundational research on cryptographic technologies and DLT scalability in general | • [44]<br>• [71]<br>• [84, 93] |
| Secure data sharing and marketplace for AI | | |
| Smart contracts | • Increase system scalability and data confidentiality<br>• Deployment and evaluation in practice, especially in permissioned ledgers within trusted consortia | • [72, 73]<br>• [44] |
| TEEs | • Deployment and evaluation of systems in practice<br>• Security analysis of TEEs, and development of secure TEEs<br>• Incorporation of incentive mechanisms (e.g., staking) | • [11, 44]<br>• [89, 94]<br>• [73] |
| Federated learning | • Deployment and evaluation in trusted consortia<br>• Security analysis and combination with other privacy techniques<br>• Analysis of incentive mechanisms (e.g., staking) | • [44]<br>• [53, 95]<br>• [73] |
| Explainable AI | | |
| - | • Deploy systems that use DLT as a trail for XAI<br>• Evaluate the practical XAI benefits of such systems | • [11, 52, 53, 69, 96]<br>• [11, 44, 53] |
| Coordination of devices | | |
| - | • Reputation management of DLT participants<br>• Self-identifying and secure devices | • [74]<br>• [75, 97] |

From our point of view, TEEs generally provide a promising trade-off between computational performance and confidentiality. Prior research that provides an ML pipeline with TEE-enabled DLTs does not evaluate its performance [11, 44, 69]. As such, a natural extension would be a practical performance evaluation. This could help researchers aiming to deploy such systems, for example, for privacy-preserving personalization. By knowing what amounts of data and what complexity of computation the system can handle, other researchers could select use cases and deploy and evaluate TEE-enabled DLTs for their use cases.

Another avenue for future research is the further development of computationally powerful TEEs, and to potentially even enable TEEs on GPUs [102], which are, for many ML model training tasks, better suited than CPUs. At the same time, research has identified security vulnerabilities in TEEs and potential security measures [94]. Therefore, future research could aim to develop secure TEEs and mechanisms to prevent attacks on TEEs, as well as study the feasibility and implications of TEEs on distributed ledgers. To prevent TEE security vulnerabilities, some researchers combine them with other privacy-enabling technologies, such as federated learning [44] or differential privacy [11, 88]. The further analysis of such combined methods with regards to data confidentiality guarantees, computational overhead, impact on machine learning quality, and practicality for use cases would be another promising avenue for further research.

In addition, TEEs with a strong integration into a blockchain protocol enable new mechanisms for block selection. These mechanisms use special functions of TEEs, such as the secure generation of random numbers, and are known as proof of luck [91] or proof-of-elapsed-time [92]. Future research could aim to further integrate such mechanisms into the DLT protocol and evaluate the practical benefits and system security implications

3) OFF-CHAIN COMPUTATION

Most presented DLT-based federated learning protocols do not ensure the integrity of the model training calculations. As such, we see future research potential for such systems in at least partially trusting consortia. Federated learning in general is already successfully applied in systems with a large number of users [51], therefore, future applied research could provide further promising results when combined with DLT.

A further avenue for future research is the use of game-theoretic mechanisms to incentivize honest computation, for example, for ML model training [71]. In general, there is a lot of current research that aims to scale smart contract executions outside a distributed ledger while maintaining integrity [93, 103-106]. Much of this research is at the conceptual level and does not yet deal in detail with the application of concrete computationally intensive applications, such as the training of ML models.

In this regard, the advancement of cryptographic technologies for secure computation is another avenue for future research. Homomorphic encryption, for example, enables data consumers to perform computations on encrypted data. However, this technology and other technologies (such as secure multiparty computation or zero knowledge proofs) are not yet practicable for ML model computations [11, 75, 84, 88] due to the strong computational overhead. If future research achieved breakthroughs to enable intensive computations with these cryptographic technologies, they would provide an alternative to TEEs.

*B. SECURE DATA SHARING AND MARKETPLACE FOR AI*

Articles within this field already present systems that could be further developed and deployed for practical use. Future research could, therefore, evaluate the practicality and user acceptance of such systems. For this practical deployment, health care use cases may be particularly suitable, because the deployment of ML in this context is especially challenging due to high privacy and security requirements [11, 84, 85].

Beyond the deployment and evaluation, a further avenue for future research is the rigorous security analysis of such systems against different types of adversaries. Potential adversaries could, for example, aim to get rewards for sharing spam data, aim to extract training data by training a proprietary ML model, or aim to extract training data of a shared ML model through inference attacks or the shared ML model itself. Only some articles of the extant literature study such attacks and security measures to protect the system [11, 53, 69]. Especially staking mechanisms have seen little study so far [73]. Future research could possibly transfer insights from proof-of-stake DLT protocols [107] toward staking in DLT-based data marketplaces.

In general, we see differential privacy as a promising technique for ensuring users' privacy [42, 70], especially in large datasets with low dimensionality. In such cases, its negative effect on the aggregated model quality is small while improving the individual data sharer's privacy. Previous research on differential privacy for ML studies its theoretical implications on test data sets [16]. In our view, future research could build on these and study differential privacy's practicality in certain vertical use cases for both, classic ML (relevant in the context of DLT; e.g., through TEEs), and federated learning. Federated learning, for example, can be vulnerable to inference attacks [86]. Differential privacy can help to prevent the practical feasibility of such attacks [95].

Articles describing DLT-based federated learning use DLT as an immutable trail and, in some cases, even for the data communication and storage. Furthermore, DLT serves as a ledger for the reward payment using cryptocurrencies. We, therefore, see plain federated learning approaches as particularly promising for applications in (at least partially) trusting consortia. TEE-based approaches, however, not only use DLT for the same reasons, but go further. Smart contracts on the distributed ledger can enforce policies for both, data providers and data consumers. From this point of view, the TEE-based approach is more powerful in terms of covering a large share of the ML pipeline, and ultimately, in ensuring system security. Accordingly, it may have better utility for future research that aims to deploy DLT-based secure data sharing and marketplace systems.

However, TEE implementations have a rich history of security vulnerabilities [108]. A detailed security analysis of the practical security of TEEs and their implications for TEE-based DLT systems, potentially with further security mechanisms in smart contracts could provide a valuable contribution to future research. Such an analysis could also include combined security measures in smart contracts, such as federated learning [44], differential privacy [11], or secure multiparty computation [11]. In general, we see such combinations of privacy techniques [65, 84] with TEEs as an interesting avenue for future research. However, some of these cryptography-based privacy techniques need breakthroughs in practicability to be usable for ML [84].

*C. EXPLAINABLE AI*
In our view, DLT is well-suited as a trail for AI model metadata and hashes of data generated around the AI model training and inference phase. As such, DLT can help to enable XAI applications that aim to describe the black box behavior of AI models for both, off-chain, as well as on-chain [11] applications. This can increase an AI system's security and, ultimately, enable the deployment of trustworthy AI [52, 96].

In many DLT-based federated learning articles, only AI model hashes and metadata [52] or relatively small AI models [53] are stored on the distributed ledger. TEE-based systems, on the other hand, can securely write and read external memory while preserving data confidentiality [69]. As such, we see TEE-based systems [11, 44, 69] as specifically suited for large, data intensive ML environments. Future research could aim to deploy such systems and evaluate the practical explainability of the AI model.

*D. COORDINATION OF DEVICES*
One of the core functions of DLT is to provide an immutable ledger where only cryptographically legitimate participants can perform transactions. As such, the usage of DLT for coordinating devices and participants is essential and basic at the same time. Reputation management of participants on a distributed ledger is one field that has only started to see attention and appears to be of potential for future research. Kang, et al. [74] specifically mention a dynamic system with variable thresholds as a future research opportunity. With regard to self-identifying TEEs, prior research already uses DLT to coordinate devices [75, 109]. We consider the further dissemination and use of self-identifying functionalities through physically unclonable functions [97] as an interesting future research opportunity. Such self-identification functionalities could potentially be built in Internet of Things devices, such as sensors or actuators. However, mechanisms for hardware-based authentication have been identified as insecure in the past. Therefore, we see an analysis of their practical security guarantees as another open research problem [110, 111].

**VIII. DISCUSSION**

*A. PRINCIPAL FINDINGS*
Both fields, AI and DLT, are currently experiencing a lot of hype within research and practice. Even though our initial database query listed a few articles that did not clearly explain what AI or especially what DLT was used for, a substantial amount of extant articles provided profound insights for our review and future research agenda.

One analysis aspect in our work is that we do not limit our work to blockchain as the only DLT concept, but also consider other concepts such as directed acyclic graphs. This decision was based on the expectation that other DLT concepts with distinct characteristics might be better suited for some AI applications than the concept of blockchain [3, 12]. However, only one out of 32 articles within our review considered DLT concepts other than blockchain in the context of AI [72]. This result is in line with previous research which has called for more research on other DLT concepts [3]. Improving and deploying DLT concepts other than blockchain is, therefore, highly interesting and a relevant avenue for future research.

In general, the classification scheme provided by Dinh and Thai [13] served as a helpful tool to classify extant literature with regard to the convergence of AI and DLT. However, several articles cover aspects from multiple groups in the subsections of AI for DLT or DLT for AI. This is particularly the case for articles in our review that cover multiple subsections of DLT for AI or the subsection privacy-preserving personalization from the section AI for DLT. We have slightly modified the scheme from Dinh and Thai [13] in two ways. First, extant literature applies AI for DLT in the context of secure distributed ledgers, but not for scalable distributed ledgers as well. A possible reason for this is the lack of AI's robustness and security guarantees, as discussed in our future research agenda. Second, we renamed the subsection *coordination of untrusting devices* to *coordination of devices*, because DLT is not the element that establishes trust in all of the articles. For example, TEEs can use physically unclonable functions for remote attestation [87].

In our view, some of the research fields are mature enough to transfer systems into practice and evaluate their influence and user acceptance. This includes, for example, the fields of security analysis of smart contracts or marketplace systems based on TEE-enabled DLTs. However, other research fields, such as an AI-based automated referee and governance for DLT protocols, appear to require substantial progress in fundamental research (e.g., robust and secure AI), before transferring scientific knowledge into practice by establishing real-world systems.

By applying convergence as a theoretical lens on extant literature, we were able to focus our research on innovative articles that closely integrate AI and DLT. Furthermore, we were able to exclude research that does not closely integrate AI and DLT, such as AI-based cryptocurrency trading or price prediction. Drawing on the definition of convergence [31], many extant articles on AI and DLT's convergence fall into the first phase with cross-scientific research on their integration. Some articles already pave the way for the second phase with new platforms arising that could, for example, accelerate health care research [11, 44, 84]. In our research, we noticed that the concept of convergence has received relatively little attention by IT researchers in the past. This came to our surprise, as convergence has been a main driver of IT innovations over the recent years [14]. Therefore, we consider convergence as a promising theoretical lens to explore interdisciplinary technological settings.

*B. LIMITATIONS*

Both fields, AI and DLT, are moving very fast and breakthroughs are regularly achieved. In this respect, we cannot rule out the possibility that in some subfields of the convergence of AI and DLT, future research will achieve innovative breakthroughs that may enable use cases not identified in our review or future research agenda. We have taken several steps to minimize the chances of this outcome. First, our analysis includes ArXiv preprints of research articles that may only soon be presented in journals or at conferences. Second, we included insights from foundational research on the topics of AI, DLT, and secure computation into our future research agenda. In doing so, we sought to consider aspects that may not be covered in the extant literature of our review today but are highly relevant in the individual fields (e.g., the practical security guarantees of TEEs). Third, we have also included articles that cover little researched technologies, often with little practicality today, but that may see breakthroughs in the future. This includes Artificial General Intelligence, DLT which is not based on the concept of blockchain, and cryptographic protocols for computationally intensive tasks.

IX. CONCLUSIONS

In this research, we investigated the current research status and future research opportunities of the convergence of AI and DLT. In order to assess the current state of convergence, we conducted a systematic literature review and analyzed extant literature through the lens of convergence. Our findings include several different ways that describe how AI can advance DLT applications, or DLT can advance AI applications. In order to develop a future research agenda, we built on the structure of our literature review and linked the ongoing research with other extant research in the separate fields of AI and DLT, as well as our own view on future research opportunities. Our results reveal multiple future research opportunities in this interdisciplinary field for both, theory- as well as practice-oriented research. With our article, we contribute to the current state of research in four ways. First, we expand prior research, which did not consider DLT concepts other than blockchain in the integration with AI. Second, we consider both perspectives, AI for DLT, and DLT for AI and the many different concepts of integration. Third, we bridge the gap between theory and practice by drawing theoretical conclusions from practical research and outlining future practical research opportunities from theory. Fourth, we describe how convergence creates innovation in an emerging field.

This article provides insights for researchers and practitioners interested in deepening their knowledge for interdisciplinary applications of any of the fields: AI, DLT, their convergence, convergence in general. By providing these insights with an overview on the upcoming convergence of DLT and AI, we contribute to the development of future innovations in this fast-paced field.